\documentclass[a4paper]{jpconf}
\usepackage{graphicx}
\usepackage{amsmath}
\usepackage{lineno}
\begin{document}
\title{Rare decays at LHCb}

\author{Marco Santimaria\footnote{on behalf of the LHCb collaboration}}

\address{INFN - Laboratori Nazionali di Frascati, via Enrico Fermi 54, 00044 Frascati (IT)}

\ead{marco.santimaria@cern.ch}

\begin{abstract}
Flavour-changing neutral-current processes, such as $b\to s \ell^+\ell^-$, are forbidden at tree level in the Standard Model and hence might receive comparatively large corrections from new particles. This document highlights recent measurements from LHCb on $b\to s \ell^+\ell^-$ and purely-leptonic decays, including tests of lepton flavour universality and searches for lepton flavour violation.
\end{abstract}

\section{Theoretical framework}
\label{sec:introduction}
Rare $b-$hadron decays offer a rich phenomenology for indirect searches for New Physics (NP), making it possible to probe energies above the $\rm{TeV}$ scale.

$b\to s \ell^+\ell^-$ decays are flavour-changing neutral-current (FCNC) processes that can only occur at the loop level in the Standard Model (SM)~\cite{Glashow:1970gm}. The presence of new particles can be spotted via precision measurements of theoretically well-known observables, such as $b \to s\mu^+\mu^-$ differential branching fractions, angular coefficients as well as ratios of $b\to s \ell^+\ell^-$ rates for different final state leptons.

Due to the non-perturbative nature of QCD interactions at the $b-$quark mass scale, an effective Hamiltonian can be built to factorise the high- and low-energy contributions in the limit $M_b \ll M_W$:
\begin{equation}
\label{eq:ha}
\mathcal{H}_{eff} = \frac{G_F}{\sqrt{2}} \sum_i V_{\rm{CKM}}^i C_i(\lambda) \mathcal{O}_i(\lambda), 
\end{equation}
where $G_F$ is the Fermi constant and $V_{\rm{CKM}}$ are the relevant CKM matrix elements. The long-distance (i.e. low-energy) contributions are described by the local operators $\mathcal{O}_i$, which are factorised from the short-distance contributions, encoded in the Wilson coefficients $C_i$, at the energy scale $\lambda$.
The SM local operators describing $b\to s\ell^+\ell^-$ processes mentioned in this document are the vector $\mathcal{O}_9^{(')}=(\overline{s}P_{L(R)}b)(\overline{l}\gamma^{\mu} l)$ and axial-vector $\mathcal{O}_{10}^{(')}=(\overline{s}P_{L(R)}b)(\overline{l}\gamma^{\mu}\gamma^5 l)$ operators, where the prime symbol represents chirality-flipped cases, as projected by $P_{L(R)}=(1\pm\gamma^5)/2$.

A generic NP contribution at the scale $\Lambda_{\rm{NP}}$ takes the form $\Delta\mathcal{H}_{eff} = (c_i/\Lambda^2_{\text{NP}})\mathcal{O}_i$, and can both alter the value of SM Wilson coefficients or introduce new operators.

\section{Branching fractions}
\begin{figure}[ht!]
\begin{center}
\includegraphics[width=.32\textwidth]{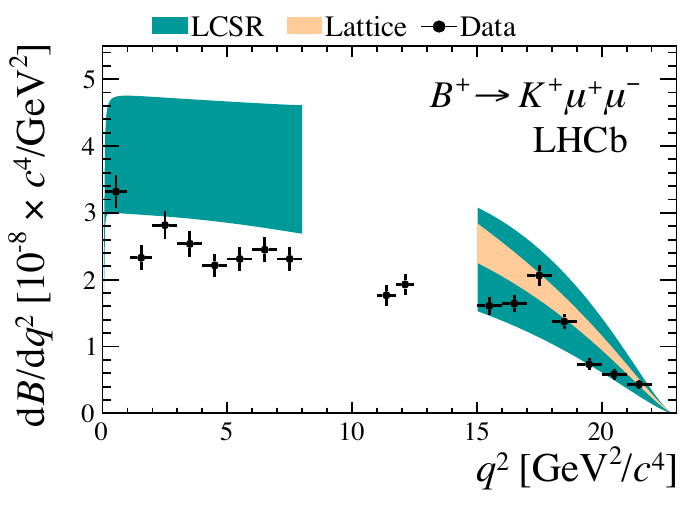}
\includegraphics[width=.32\textwidth]{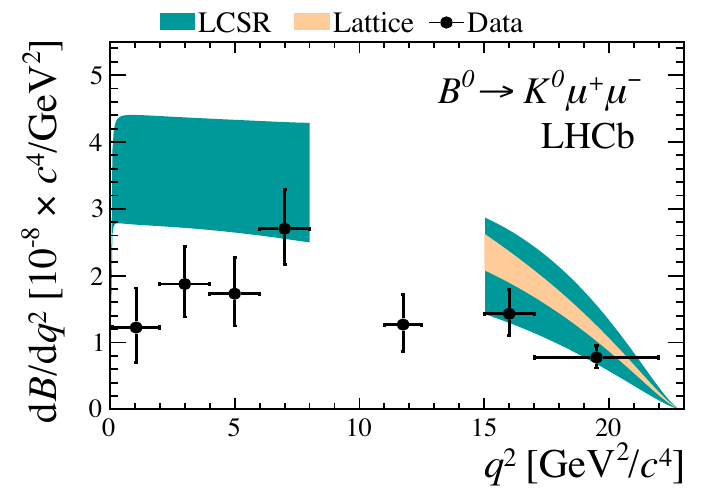}
\includegraphics[width=.32\textwidth]{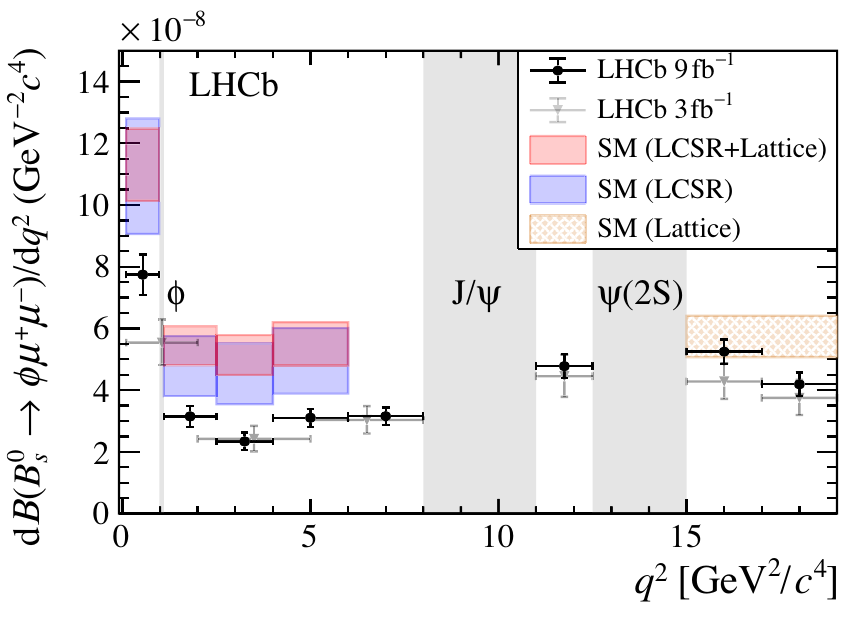}
\end{center}
\caption{\label{fig:rates} Differential branching fractions of $B^+ \to K^+ \mu^+\mu^-$, $B^0 \to K^0 \mu^+\mu^-$ and $B^0_s \to \phi \mu^+\mu^-$ decays measured at LHCb, compared with LCSR and lattice form factor predictions~\cite{LHCb:2014cxe, LHCb:2021zwz}.}
\end{figure}

Differential branching fractions of various $b\to s\mu^+\mu^-$ decays have been measured at LHCb as a function of the dilepton invariant mass squared, $q^2$. Some examples are shown in Fig.~\ref{fig:rates}.

In both Run 1 and Run 2 data, measured rates appear to be consistently lower with respect to the SM predictions, whose uncertainty is dominated by the form factors. On the experimental side, the precision is limited by the knowledge of $B \to J/\psi X$ branching fractions used for normalisation. For these reasons, the discrepancies cannot be attributed to anomalous muonic couplings: observables with reduced hadronic contributions should be investigated.

\subsection{Purely-leptonic decays}
\label{ssec:leptonic}
\begin{figure}[ht!]
\begin{center}
\includegraphics[width=.44\textwidth]{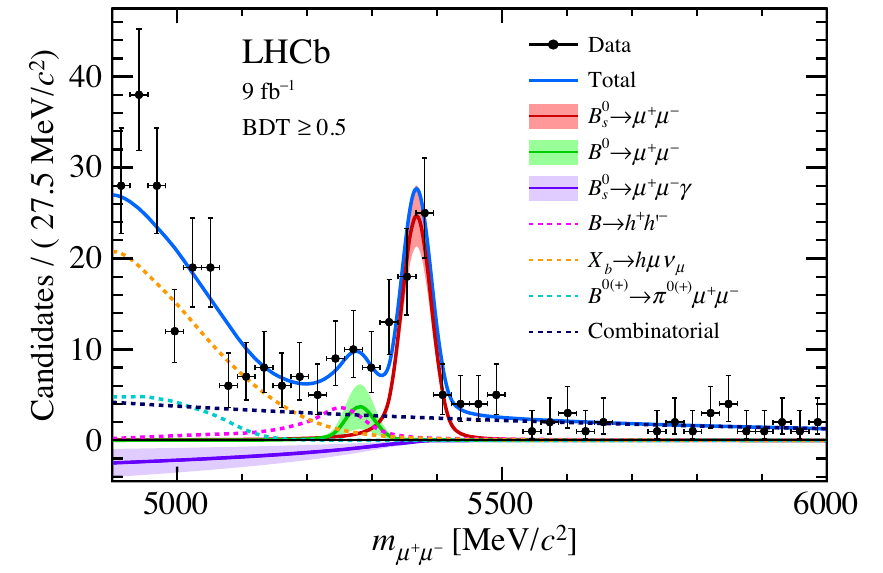}
\hfill
\includegraphics[width=.49\textwidth]{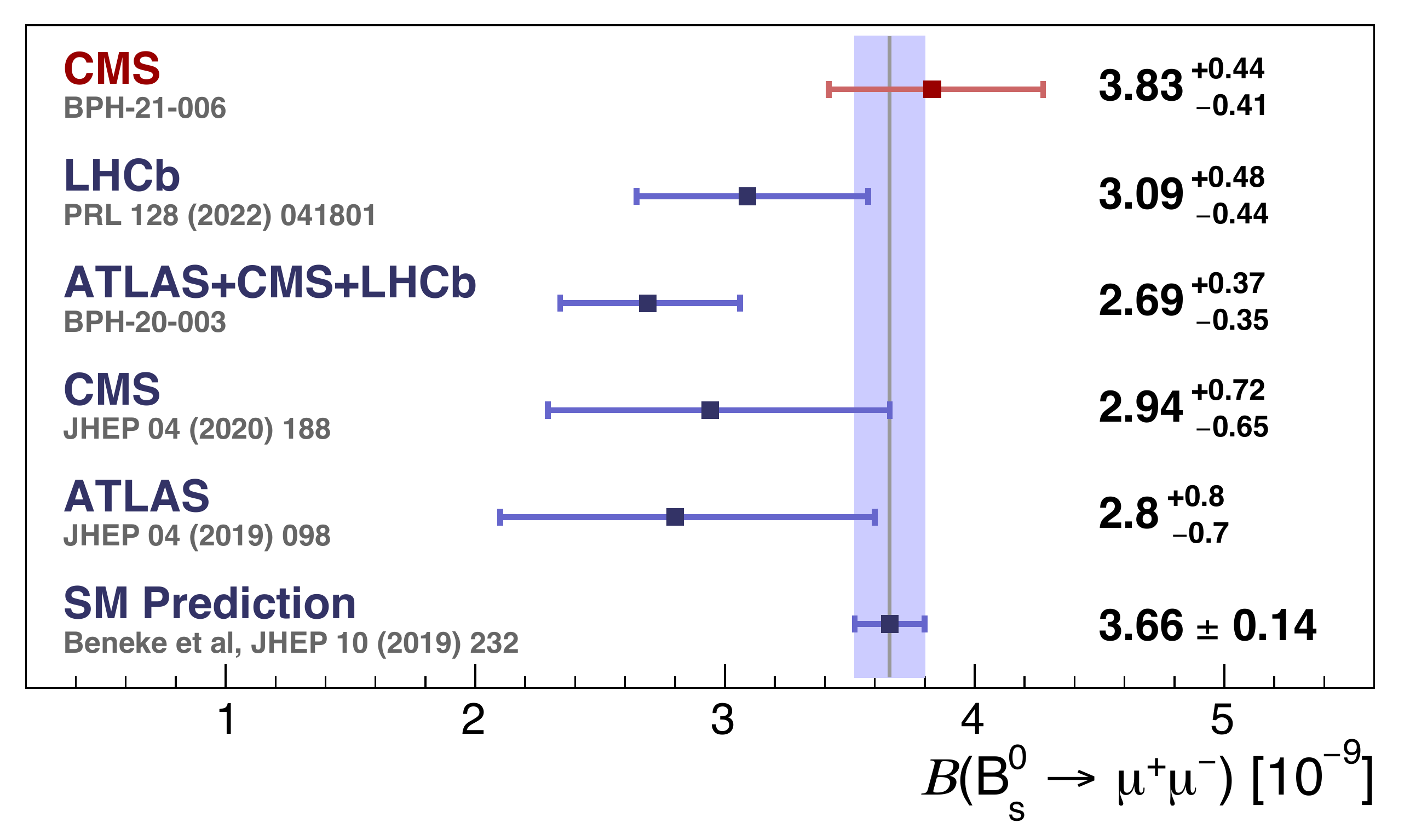}
\end{center}
\caption{\label{fig:bsmumu} Left: Dimuon invariant mass spectrum for the LHCb analysis of $B^0_{(s)} \to \mu^+\mu^-$ decays. Right: measurements of the $B^0_s \to \mu^+\mu^-$ branching fraction from LHC experiments compared to the SM prediction.}
\end{figure}

Purely-leptonic final states allow for more precise theoretical predictions. For example, the SM branching fraction of $B^0_s \to \mu^+\mu^-$ decays depends only on the Wilson coefficient $C_{10}$ and on a single hadronic constant, known at $\approx 0.5\%$~\cite{Bazavov:2017lyh}, leading to the precise prediction $\mathcal{B}(B^0_s\to\mu^+\mu^-) = (3.66\pm 0.14)\times10^{-9}$~\cite{Beneke:2019slt}.
The most-precise LHCb measurement, shown in Fig.~\ref{fig:bsmumu} (left), agrees with this prediction within one standard deviation~\cite{LHCb:2021vsc,LHCb:2021awg}.
A recent precise measurement from the CMS experiment confirms SM compatibility~\cite{CMS:2022dbz}, and is reported in Fig.~\ref{fig:bsmumu} (right) together with the latest LHC results.

Other fully-leptonic decays are analysed at LHCb, leading to the world's best upper limits summarised in Tab.~\ref{tab:leptonic}.

\begin{center}
\begin{table}[h]
\centering
\caption{\label{tab:leptonic} Upper limits on the branching fraction of various fully-leptonic decays at LHCb. A light scalar with $m(a)=1~\rm{GeV}$ is assumed.} 
\begin{tabular}{@{}l*{15}{l}}
\br
Channel&$\mathcal{B}$ upper limit (95\% CL)&Reference\\
\mr
$B^0_{(s)} \to e^+e^-$& $3.0~(11.2)\times10^{-9}$ &\cite{LHCb:2020pcv}\\
$B^0_{(s)} \to \tau^+\tau^-$& $2.1~(6.8)\times10^{-3}$ &\cite{LHCb:2017myy}\\
$B^0_{(s)} \to \mu^+\mu^-\mu^+\mu^-$ & $1.8~(8.6)\times10^{-10}$ &\cite{LHCb:2021iwr}\\
$B^0_{(s)} \to a(\mu^+\mu^-)a(\mu^+\mu^-)$ & $2.3~(5.8)\times10^{-10}$ &\cite{LHCb:2021iwr}\\
$B^0_{(s)} \to J/\psi(\mu^+\mu^-)\mu^+\mu^-$ & $1.0~(2.6)\times10^{-9}$ &\cite{LHCb:2021iwr}\\
\br
\end{tabular}
\end{table}
\end{center}

\section{Angular observables}
\begin{figure}[ht!]
\begin{center}
\includegraphics[width=.55\textwidth]{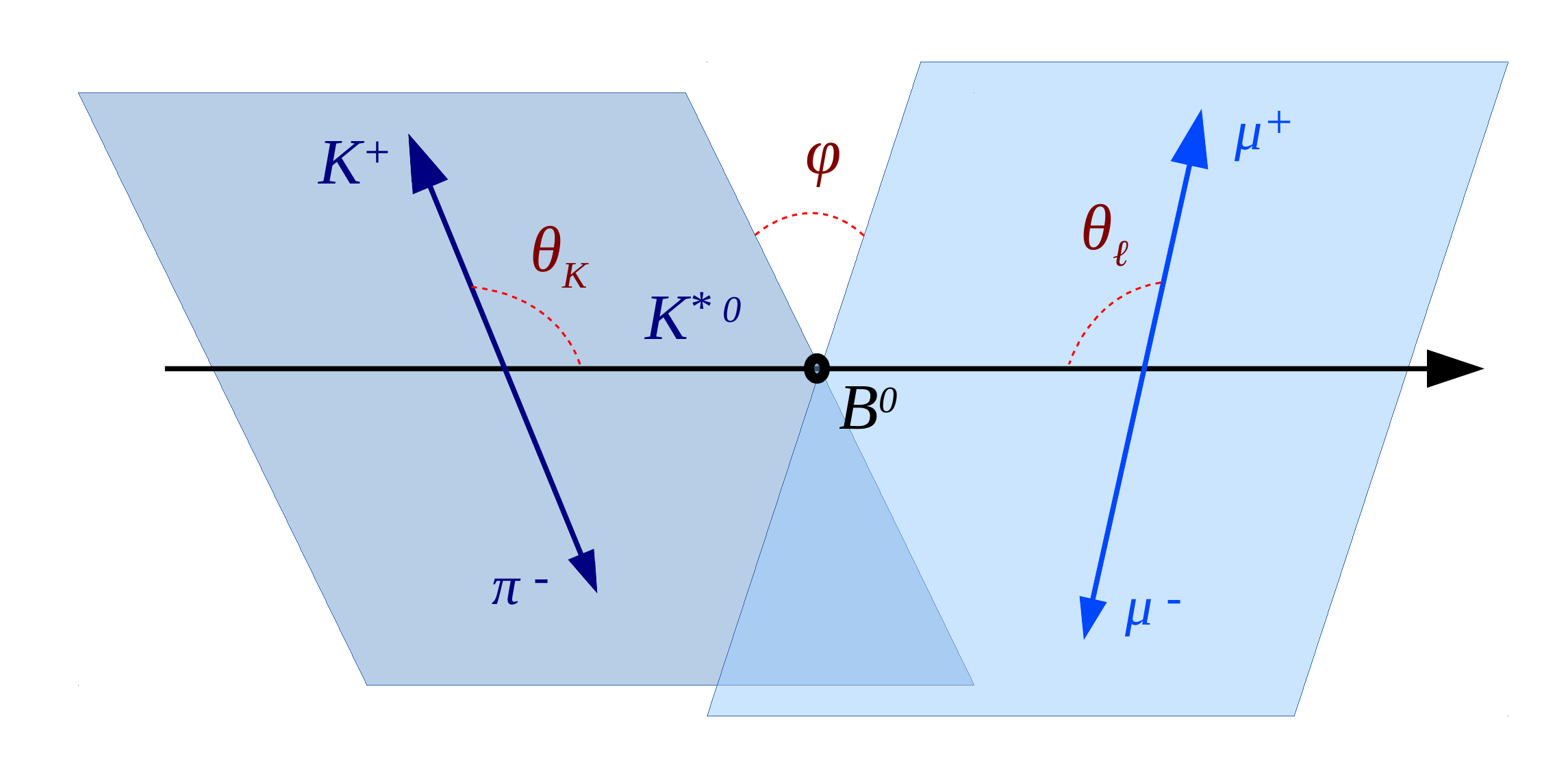}
\hfill
\includegraphics[width=.42\textwidth]{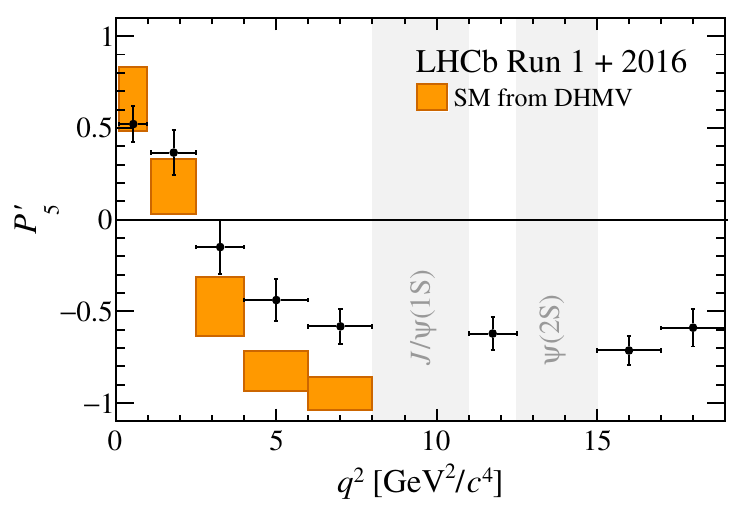}
\end{center}
\caption{\label{fig:ang} Left: angular basis from the $B^0 \to K^{*0}\mu^+\mu^-$ analysis. Right: $P_5'$ measurement from LHCb compared with the SM prediction~\cite{LHCb:2020lmf}.}
\end{figure}

The differential decay rate of a $B$ meson to a vector meson (e.g. a $K^{*0}$) and two leptons can be described by the $q^2$ and the three angles $\overrightarrow{\Omega} = (\phi,\theta_K,\theta_L)$, defined in Fig.~\ref{fig:ang} (left):
\begin{equation}
\label{eq:ang}
\frac{d^4\overline{\Gamma}[B^0 \to K^{*0}\mu^+\mu^-]}{dq^2 d\overrightarrow{\Omega}} = \frac{9}{32\pi}\sum_i \overline{I}_i(q^2) f_i(\overrightarrow{\Omega}),
\end{equation}
i.e. a superposition of angular moments $f_i$ with coefficients $I_i$.

The leading hadronic uncertainty cancels in the combination of angular coefficients
\begin{equation}
    P_5' = \frac{S_5}{\sqrt{F_L(1-F_L)}},
\end{equation}
where $S_5$ is a CP-averaged coefficient and $F_L$ represents the longitudinal polarisation fraction of the $K^{*0}$~\cite{Matias:2012xw}.

The most-precise LHCb measurement of $P_5'$ is reported in Fig.~\ref{fig:ang} (right) from the angular analysis of $B^0 \to K^{*0} \mu^+\mu^-$ decays performed on Run 1 and a portion of Run 2 data~\cite{LHCb:2020lmf}. The results show local deviations in the fourth and fifth $q^2$ bins of $2.5~\sigma$ and $2.9~\sigma$ with respect to the SM prediction. The data can be explained with a $\sim3~\sigma$ shift on the real part of the Wilson coefficient $C_9$, although this deviation depends on the choice of the SM nuisance parameters such as form factors and long-distance contributions from the charmonium modes.

A similar trend in $P_5'$ is observed in the LHCb analysis of $B^+ \to K^{*+}\mu^+\mu^-$ decays~\cite{LHCb:2020gog}, whereas a recent measurement of the observable $F_L$ on $B^0_s \to \phi \mu^+\mu^-$ decays agrees with the SM prediction~\cite{LHCb:2021xxq}.

\section{Ratios of branching fractions}
Lepton Flavour Universality (LFU), i.e. the fact that lepton electroweak couplings are equal, is a property of the SM which has been verified in several channels~\cite{Workman:2022ynf}. However, the discrepancies highlighted in the previous sections prompted the verification of LFU in $b-$hadron decays.
Penguin decays of $b-$hadrons can serve as LFU tests via measurements of the ratios of branching fractions:
\begin{equation}
    R_H = \frac{\int^{q^2_{max}}_{q^2_{min}}\frac{d\Gamma [B\to H \mu^+\mu^-]}{dq^2}dq^2}{\int^{q^2_{max}}_{q^2_{min}}\frac{d\Gamma [B\to H e^+e^-]}{dq^2}dq^2},
\end{equation}
with $H$ being a strange hadron. These observables are free from QCD contributions and are predicted in the SM with percent precision due to high-order QED effects~\cite{Bordone:2016gaq}.

Since LFU in $J/\psi \to \ell^+ \ell^-$ decays is verified with per mille precision~\cite{Workman:2022ynf}, the LHCb observables are built as double ratios with respect to the $J/\psi$ resonant modes, in order to profit from the cancellation of detection differences between electrons and muons. 
In the case of $B^+ \to K^+ \ell^+ \ell^-$ decays, the ratio is defined as:
\begin{equation}
    R_K = \frac{\mathcal{B}(B^+ \to K^+ \mu^+\mu^-)}{\mathcal{B}(B^+ \to K^+ e^+e^-)} \bigg/ \frac{\mathcal{B}(B^+ \to K^+ J/\psi(\mu^+\mu^-))}{\mathcal{B}(B^+ \to K^+ J/\psi(e^+e^-))},
\end{equation}
where resonant and rare modes are distinguished by their $q^2$ values.

\begin{figure}[ht!]
\begin{center}
\includegraphics[width=.49\textwidth]{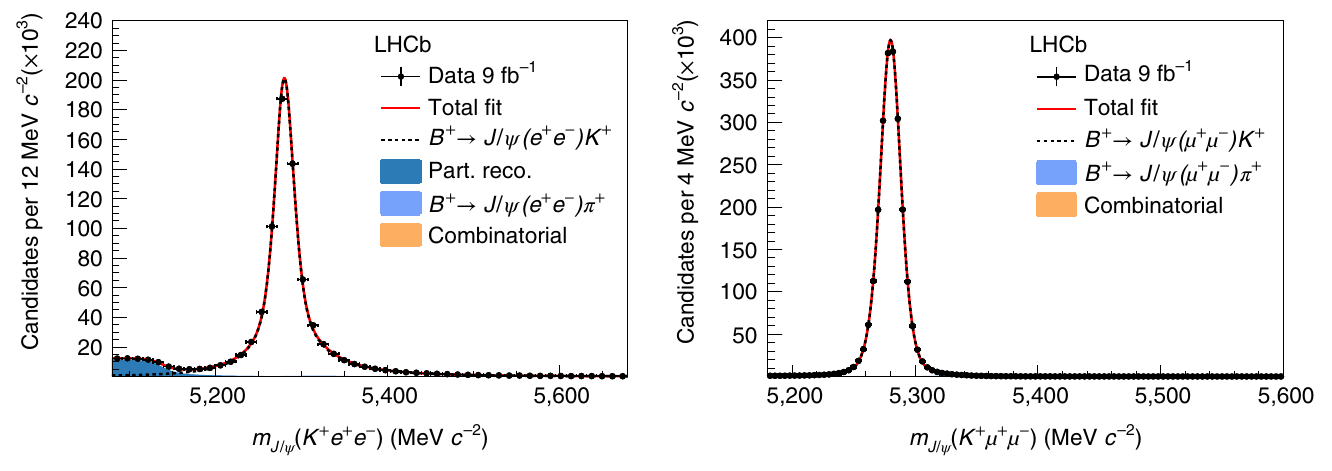}
\hfill
\includegraphics[width=.49\textwidth]{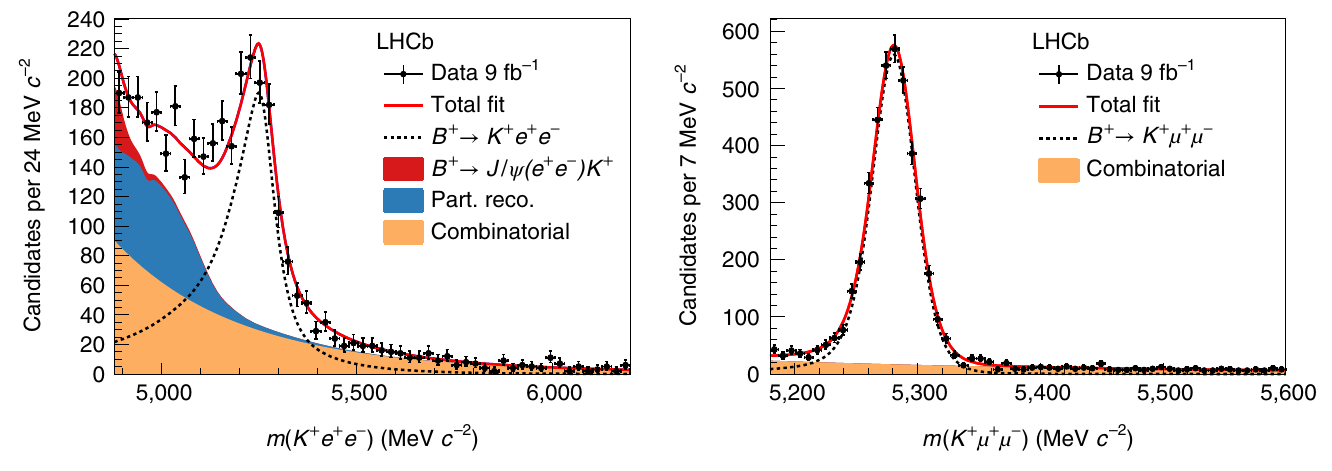}
\end{center}
\caption{\label{fig:lfu} Mass spectra of $B^+ \to J/\psi(\ell^+\ell^-) K^+$ (left) and $B^+ \to K^+ \ell^+\ell^-$ (right) decays~\cite{LHCb:2021trn}.}
\end{figure}

Fig.~\ref{fig:lfu} shows the mass spectra of the electron and muon modes from the latest $R_K$ measurement at LHCb~\cite{LHCb:2021trn}, yielding the most precise LFU test in this sector.
Several other LFU tests have been conducted at LHCb on different decay modes and are summarised in Tab.~\ref{tab:lfu}. 
\begin{center}
\begin{table}[h]
\centering
\caption{\label{tab:lfu} LFU measurements at LHCb. The first uncertainty is statistical and the second systematic.} 
\begin{tabular}{@{}l*{15}{l}}
\br
Channel&$q^2$ range&$R_H$ value&SM compatibility&Reference\\
\mr
$B^+ \to K^+ \ell^+\ell^-$& $1.1-6.0~\rm{GeV^2}$ & $R_K = 0.846^{+0.042+0.013}_{-0.039-0.012}$ &$3.1~\sigma$&\cite{LHCb:2021trn}\\

$B^0 \to K^{*0} \ell^+\ell^-$& $0.045-1.1~\rm{GeV^2}$ &$R_{K^{*0}} = 0.66^{+0.11}_{-0.07} \pm 0.03$ &$2.1~\sigma$&\cite{LHCb:2017avl}\\

$B^0 \to K^{*0} \ell^+\ell^-$& $1.1-6.0~\rm{GeV^2}$ &$R_{K^{*0}} = 0.69^{+0.11}_{-0.07} \pm 0.05$ &$2.4~\sigma$&\cite{LHCb:2017avl}\\

$\Lambda^0_b \to pK^- \ell^+\ell^-$& $0.1-6.0~\rm{GeV^2}$ &$R_{pK} = 0.86^{+0.14}_{-0.11} \pm 0.05$ & $<1~\sigma$&\cite{LHCb:2019efc}\\

$B^0 \to K^0_s \ell^+\ell^-$& $1.1-6.0~\rm{GeV^2}$ & $R_{K^0_s} = 0.66^{+0.20+0.02}_{-0.14-0.04}$ &$1.5~\sigma$&\cite{LHCb:2021lvy}\\

$B^+ \to K^{*+} \ell^+\ell^-$& $0.045-6.0~\rm{GeV^2}$ & $R_{K^{*+}} = 0.70^{+0.18+0.03}_{-0.13-0.04}$ &$1.4~\sigma$&\cite{LHCb:2021lvy}\\

\br
\end{tabular}
\end{table}
\end{center}

\section{Searches for Lepton Flavour Violating decays}
The conservation of the lepton flavour is an accidental symmetry of the SM, known to be broken in the neutrino sector. However, the induced charged Lepton Flavour Violation (LFV) occurs at immeasurably small rates of order $10^{-54}$~\cite{Calibbi:2017uvl}.
NP models contemplating a violation of the LFU, e.g. via Z' or Leptoquark mediators, typically do not conserve lepton flavour, and can thus enhance LFV decay rates to experimentally accessible levels~\cite{Glashow:2014iga}.
Numerous searches have been conducted at LHCb, leading to the world's best limits on the $B$ decay modes reported in Tab.~\ref{tab:lfv}.

\begin{center}
\begin{table}[h]
\centering
\caption{\label{tab:lfv} Upper limits on the branching fraction of LFV modes searched at LHCb.} 
\begin{tabular}{@{}l*{15}{l}}
\br
Channel&$\mathcal{B}$ upper limit (95\% CL)&Reference\\
\mr
$B^0_{(s)} \to e^{\pm}\mu^{\mp}$ & $1.3(6.3)\times10^{-9}$ & \cite{LHCb:2017hag} \\
$B^0_{(s)} \to \tau^{\pm}\mu^{\mp}$ & $1.4(4.2)\times10^{-5}$ & \cite{LHCb:2019ujz} \\
$B^+ \to K^+\mu^-e^+$ & $9.5\times10^{-9}$ & \cite{LHCb:2019bix} \\
$B^+ \to K^+\mu^+e^-$ & $8.8\times10^{-9}$ & \cite{LHCb:2019bix} \\
$B^+ \to K^+\mu^-\tau^+$ & $5\times10^{-5}$ & \cite{LHCb:2020khb} \\
$B^0 \to K^{*0}\mu^{\pm}e^{\mp}$ & $11.7\times10^{-9}$ & \cite{LHCb:2022lrd} \\
$B^0_s \to \phi\mu^{\pm}e^{\mp}$ & $19.8\times10^{-9}$ & \cite{LHCb:2022lrd} \\
$B^0 \to K^{*0}\tau^+\mu^-$ & $1.2\times10^{-5}$ & \cite{LHCb:2022wrs} \\
$B^0 \to K^{*0}\tau^-\mu^+$ & $9.8\times10^{-6}$ & \cite{LHCb:2022wrs} \\
$B^0_{(s)} \to p\mu^- $ & $3.1(14.0)\times10^{-9}$ & \cite{Collaboration:2022mzg} \\
\br
\end{tabular}
\end{table}
\end{center}

\section{Conclusions}
In recent years, several deviations with respect to the SM predictions have been observed in $b \to s \ell^+\ell^-$ decays at LHCb. Although individual significances do not exceed $2-3$ standard deviations, a coherent pattern seem to emerge:
further experimental investigation is required to clarify these anomalies.

\section*{References}
\bibliography{bib}

\providecommand{\newblock}{}
\begin{thebibliography}{10}
\expandafter\ifx\csname url\endcsname\relax
  \def\url#1{{\tt #1}}\fi
\expandafter\ifx\csname urlprefix\endcsname\relax\def\urlprefix{URL }\fi
\providecommand{\eprint}[2][]{\url{#2}}

\bibitem{Glashow:1970gm}
Glashow S, Iliopoulos J and Maiani L 1970 {\em Phys. Rev. D\/} {\bf 2}
  1285--1292

\bibitem{LHCb:2014cxe}
Aaij R {\em et~al.\/} (LHCb) 2014 {\em JHEP\/} {\bf 06} 133 (\textit{Preprint}
  \eprint{1403.8044})

\bibitem{LHCb:2021zwz}
Aaij R {\em et~al.\/} (LHCb) 2021 {\em Phys. Rev. Lett.\/} {\bf 127} 151801
  (\textit{Preprint} \eprint{2105.14007})

\bibitem{Bazavov:2017lyh}
Bazavov A {\em et~al.\/} 2018 {\em Phys. Rev. D\/} {\bf 98} 074512
  (\textit{Preprint} \eprint{1712.09262})

\bibitem{Beneke:2019slt}
Beneke M, Bobeth C and Szafron R 2019 {\em JHEP\/} {\bf 10} 232
  (\textit{Preprint} \eprint{1908.07011})

\bibitem{LHCb:2021vsc}
Aaij R {\em et~al.\/} (LHCb) 2022 {\em Phys. Rev. Lett.\/} {\bf 128} 041801
  (\textit{Preprint} \eprint{2108.09284})

\bibitem{LHCb:2021awg}
Aaij R {\em et~al.\/} (LHCb) 2022 {\em Phys. Rev. D\/} {\bf 105} 012010
  (\textit{Preprint} \eprint{2108.09283})

\bibitem{CMS:2022dbz}
CMS 2022  \urlprefix\url{https://cds.cern.ch/record/2815334}

\bibitem{LHCb:2020pcv}
Aaij R {\em et~al.\/} (LHCb) 2020 {\em Phys. Rev. Lett.\/} {\bf 124} 211802
  (\textit{Preprint} \eprint{2003.03999})

\bibitem{LHCb:2017myy}
Aaij R {\em et~al.\/} (LHCb) 2017 {\em Phys. Rev. Lett.\/} {\bf 118} 251802
  (\textit{Preprint} \eprint{1703.02508})

\bibitem{LHCb:2021iwr}
Aaij R {\em et~al.\/} (LHCb) 2022 {\em JHEP\/} {\bf 03} 109 (\textit{Preprint}
  \eprint{2111.11339})

\bibitem{Matias:2012xw}
Matias J, Mescia F, Ramon M and Virto J 2012 {\em JHEP\/} {\bf 04} 104
  (\textit{Preprint} \eprint{1202.4266})

\bibitem{LHCb:2020lmf}
Aaij R {\em et~al.\/} (LHCb) 2020 {\em Phys. Rev. Lett.\/} {\bf 125} 011802
  (\textit{Preprint} \eprint{2003.04831})

\bibitem{LHCb:2020gog}
Aaij R {\em et~al.\/} (LHCb) 2021 {\em Phys. Rev. Lett.\/} {\bf 126} 161802
  (\textit{Preprint} \eprint{2012.13241})

\bibitem{LHCb:2021xxq}
Aaij R {\em et~al.\/} (LHCb) 2021 {\em JHEP\/} {\bf 11} 043 (\textit{Preprint}
  \eprint{2107.13428})

\bibitem{Workman:2022ynf}
Workman R~L and Others (Particle Data Group) 2022 {\em PTEP\/} {\bf 2022}
  083C01

\bibitem{Bordone:2016gaq}
Bordone M, Isidori G and Pattori A 2016 {\em Eur. Phys. J. C\/} {\bf 76} 440
  (\textit{Preprint} \eprint{1605.07633})

\bibitem{LHCb:2021trn}
Aaij R {\em et~al.\/} (LHCb) 2022 {\em Nature Phys.\/} {\bf 18} 277--282
  (\textit{Preprint} \eprint{2103.11769})

\bibitem{LHCb:2017avl}
Aaij R {\em et~al.\/} (LHCb) 2017 {\em JHEP\/} {\bf 08} 055 (\textit{Preprint}
  \eprint{1705.05802})

\bibitem{LHCb:2019efc}
Aaij R {\em et~al.\/} (LHCb) 2020 {\em JHEP\/} {\bf 05} 040 (\textit{Preprint}
  \eprint{1912.08139})

\bibitem{LHCb:2021lvy}
Aaij R {\em et~al.\/} (LHCb) 2022 {\em Phys. Rev. Lett.\/} {\bf 128} 191802
  (\textit{Preprint} \eprint{2110.09501})

\bibitem{Calibbi:2017uvl}
Calibbi L and Signorelli G 2018 {\em Riv. Nuovo Cim.\/} {\bf 41} 71--174
  (\textit{Preprint} \eprint{1709.00294})

\bibitem{Glashow:2014iga}
Glashow S~L, Guadagnoli D and Lane K 2015 {\em Phys. Rev. Lett.\/} {\bf 114}
  091801 (\textit{Preprint} \eprint{1411.0565})

\bibitem{LHCb:2017hag}
Aaij R {\em et~al.\/} (LHCb) 2018 {\em JHEP\/} {\bf 03} 078 (\textit{Preprint}
  \eprint{1710.04111})

\bibitem{LHCb:2019ujz}
Aaij R {\em et~al.\/} (LHCb) 2019 {\em Phys. Rev. Lett.\/} {\bf 123} 211801
  (\textit{Preprint} \eprint{1905.06614})

\bibitem{LHCb:2019bix}
Aaij R {\em et~al.\/} (LHCb) 2019 {\em Phys. Rev. Lett.\/} {\bf 123} 241802
  (\textit{Preprint} \eprint{1909.01010})

\bibitem{LHCb:2020khb}
Aaij R {\em et~al.\/} (LHCb) 2020 {\em JHEP\/} {\bf 06} 129 (\textit{Preprint}
  \eprint{2003.04352})

\bibitem{LHCb:2022lrd}
LHCb 2022  (\textit{Preprint} \eprint{2207.04005})

\bibitem{LHCb:2022wrs}
LHCb 2022  (\textit{Preprint} \eprint{2209.09846})

\bibitem{Collaboration:2022mzg}
LHCb 2022  (\textit{Preprint} \eprint{2210.10412})

\end{thebibliography}


\end{document}